\documentstyle[twoside,fleqn,espcrc2]{article}

\newcommand{\BF}[1]{\mbox{\boldmath $#1$}}

\newcommand{\AmS}{{\protect\the\textfont2
  A\kern-.1667em\lower.5ex\hbox{M}\kern-.125emS}}

\title{ High precision single-cluster Monte Carlo measurement of
the critical exponents of the classical 3D Heisenberg model
\thanks{
Work supported in part by Deutsche
Forschungsgemeinschaft under grant Kl256. Preprints FUB-HEP 21/92
and HLRZ 89/92.}}

\author{C. Holm\address{Institut f\"{u}r Theoretische Physik,
                      Freie Universit\"{a}t Berlin\\
        Arnimallee 14, D-1000 Berlin 33, Germany}%
        and
        W. Janke\address{H\"ochstleistungsrechenzentrum,
                      Forschungszentrum J\"ulich\\
        Postfach 1913, D-5170 J\"ulich, Germany}}
\begin{document}

\begin{abstract}
We report measurements of the critical exponents of the classical
three-dimensional Heisenberg model on simple cubic lattices of size $L^3$ with
$L$ = 12, 16, 20, 24, 32, 40, and 48. The data was obtained from a few long
single-cluster Monte Carlo simulations near the phase transition. We compute
high precision estimates of the critical coupling $K_c$, Binder's parameter
$U^*$,
and the critical
exponents $\nu,\beta / \nu,\eta$, and $\alpha / \nu$,
using extensively histogram reweighting and optimization techniques
that allow us to keep control over the statistical errors.
Measurements of the autocorrelation time show  the expected reduction of
critical slowing down at the phase transition as compared to
local update algorithms. This allows simulations on
significantly larger lattices than in previous studies and consequently a
better
control over systematic errors in finite-size scaling analyses.
\end{abstract}

\maketitle
%
%
\section{INTRODUCTION}
The three-dimensional (3D) classical Heisenberg
model is one of the simplest spin models, and its critical behavior has
been investigated
by a variety of approaches.
Recently Peczak, Ferrenberg, and Landau \cite{peczak91} (PFL)
have undertaken a high
statistics Monte Carlo
(MC) study of this model on cubic lattices of sizes up to $V = L^3 = 24^3$,
using standard Metropolis and
multi-histogram techniques \cite{fs}.
They found a value
of $K_c = 0.6929(1)$  that was in disagreement with previous estimates
of  the critical
coupling $K_c=J/k_BT_c$.
These estimates were derived from high-temperature series (HTS)
expansion analyses \cite{newseries} based on the Pad\'e
($K_c =0.6924(2)$) and ratio ($K_c =
0.6925(1)$) method, respectively, and more recent transfer-matrix (TM)
MC investigations \cite{bloete} ($K_c = 0.6922(2)$ and $K_c =
0.6925(3)$).
The critical coupling is a non-universal parameter and from this point of view
of no particular interest. Most estimates of universal critical exponents,
however, are biased and usually depend quite strongly on the precise value
of $K_c$. To clarify the above discrepancy we
performed an independent high precision single-cluster MC study on large
 lattices
of sizes up to $48^3$. We could confirm PFL's value of $K_c$ and obtained
measurements of the critical exponents that are
in their accuracy comparable to the best estimates coming from field theory.
\section{MODEL AND ALGORITHM}
The Metropolis (pseudo) dynamics
suffers from the severe problem of critical slowing down, which is the reason
why we used as update algorithm the cluster algorithm in
its single-cluster version \cite{wolff}. One update in the single-cluster
variant
consists of chosing a random mirror plane and a random starting site, which
is the
germ of a cluster whose members are selected from adjacent sites by a
Metropolis-like accept/reject criterion. From studies of related spin models it
is known \cite{othermodels}  that this variant of the cluster algorithm
is extremely efficient in three dimensions.

The classical Heisenberg model is defined by the Hamiltonian
\begin{equation}
{\cal H} = J \sum_{\langle i,j \rangle} \left[
1-\vec{s}_i \cdot \vec{s}_j \right]
\label{eq:2}
\end{equation}
where $J$ is the ferromagnetic coupling ($J>0$).
The sum runs
over all nearest neighbour pairs $\langle i,j \rangle$, and $\vec{s}$ are
three-dimensional unit spins  that live on the sites $i$ of a simple cubic
lattice with periodic boundary conditions.
The continuous energy range $0 \le E \le 3V$ was
discretized into 90000 bins, which is fine enough to
ensure no significant discretization errors.
We stored histograms instead of the whole data time
series in order to save storage space.

Our simulations were organized as follows. First,
we did one run for each lattice size at $K_0 = 0.6929$, the estimate of
$K_c$ by PFL, and recorded the energy histogram $P_{K_0}(E)$ and the
microcanonical averages
$\langle\!\langle m^k \rangle\!\rangle(E) \equiv \sum_M P_{K_0}(E,M)
m^k/P_{K_0}(E)$,
$k=1,2,4$,
where $m \equiv M/V = |\vec{m}|$ is the magnitude of the magnetization
$\vec{m} =
\frac{1}{V} \sum_{\BF{x}} \vec{s}(\BF{x})$ of a single spin configuration.
The temperature independent averages
$\langle\!\langle m^k \rangle\!\rangle(E)$
can be computed by accumulating the values of $m^k$ in lists
indexed by the associated energy bin of the configuration and normalizing
at the end by the total number of entries in each bin, making it thus
unnecessary to store the two-dimensional
histogram $P_{K_0}(E,M)$.

{}From the data of the $K_0$ run we could compute
the approximate
positions $K_+ > K_0$ and $K_- < K_0$ of the (connected)
susceptibility and the specific-heat peak maximum
by reweighting techniques \cite{fs}. We then performed two more runs at
$K_+$ and $K_-$, respectively,
again recording $P_{K}(E)$ and $\langle\!\langle m^k \rangle\!\rangle(E)$.
This choice of the simulation points has the advantage that one
automatically stays in the critical
region since both $K_+$ and $K_-$ scale with $L^{-1/\nu}$, where $\nu$
is the correlation length exponent.
{}From this data we then computed three estimates
${\cal O}_L^{(n)}(K)$, $n = -,0,+$ for all thermodynamic observables
${\cal O}_L$ of interest,
and for any $K$ value in the vicinity of
$K_-$, $K_0$, $K_+$ by reweighting.
The reweighting range was determined by the energy value at which the
 energy histogram
had decreased to a third of its maximum. This ensured a high enough statistics
of the histogram to allow the reweighting scheme to produce reliable results.
Furthermore we used blocking to compute
jackknife errors $\Delta {\cal O}_L^{(n)}$ on ${\cal O}_L^{(n)}$.
To get an optimized average of these three values that minimizes the
relative error of the combined ${\cal O}_L(K)$ for each observable
separately, we added the ${\cal O}_L^{(n)}$ with
relative weight $1/(\Delta {\cal O}_L^{(n)})^2$.
All our runs contain at least $10000 \times \tau$ measurements, where $\tau$
is the integrated autocorrelation time of the susceptibility. The measured
value of $\tau$ turned out to be almost independent of the
lattice size and to be very small, $\tau <2$, in units of
lattice sweeps that allow direct comparison with the Metropolis algorithm.
This is of course expected
for the single-cluster update.
For the $48^3$ lattice we obtain a value of $\tau$ which is about
three orders of magnitude smaller than for the Metropolis algorithm. This
explains why we could study much larger lattice sizes than PFL, and could
still afford to have about ten times better statistics.
\section{RESULTS}
To determine $K_c$ we first concentrated on Binder's parameter
\begin{equation}
U_L(K) = 1 - \frac{\langle m^4 \rangle}
{3 \langle m^2 \rangle^2}.
\label{eq:3}
\end{equation}
Asymptotically for large $L$, all curves
$U_L(K)$ should cross in the unique point  $(K_c,U^*)$.
The locations of the crossing points of two
different curves $U_L(K)$ and $U_{L'}(K)$ depend on the scale
factor $b=L/L'$, due to residual
corrections to finite-size scaling (FSS). From the crossings with the $L=12$
and $L=16$
curves we obtained six and five data points, respectively, whose linear
 fits shown
in Fig.~1 gave us the result
\begin{equation}
K_c =0.6930 \pm 0.0001,
\label{eq:4}
\end{equation}
and
\begin{equation}
U^* = 0.6217 \pm 0.0008.
\label{eq:5}
\end{equation}
The critical coupling is found in excellent agreement with the value
quoted by PFL, and also $U^*$ agrees very well with their estimate of
$0.622(1)$. For comparison, a field theoretic $\epsilon$-expansion
predicts a $4\%$ lower value
of $0.59684...$ \cite{brezin}.

To obtain an estimate for the correlation length exponent
$\nu$ we use that at $K_c$
the derivatives $dU_L/dK$
should scale asymptotically with $L^{1/\nu}$.
To calculate $dU_L/dK$ from our data, we took the
thermodynamic derivative
\begin{equation}
{{dU_L} \over {dK }}=(1-U)\left\{ {\langle E\rangle -2{{\langle m^2E\rangle
}  \over {\left\langle {m^2} \right\rangle}}+{{\langle m^4E\rangle } \over
{\left\langle {m^4} \right\rangle}}} \right\}, \label{eq:18a}
\end{equation}
which is less vulnerable to systematic errors than a finite difference
approximation scheme. In a log-log
plot we find a perfect straight line fit (with
goodness-of-fit parameter $Q=0.61$ at $K_c= 0.6930$) that yields
\begin{equation}
\nu = 0.704 \pm 0.006.
\label{eq:6}
\end{equation}
This value is in good agreement with PFL's measurement of
$\nu = 0.706(9)$ (determined by the same method, but at $K = 0.6929$), and
with the field theoretic estimates derived
from the resummed $g$-expansion \cite{pert} or from the
resummed $\epsilon$-expansion \cite{epsilon}; see Table~1.
The high quality of this fit (as well
as of all other fits described below) shows that the asymptotic scaling
formula works down to our smallest lattice size $L=12$, indicating that
there is no need for confluent correction terms.

Having now measured $\nu$ we can get two more estimates for
the critical coupling by assuming the FSS relation
$T_{\rm max} = T_c + a L^{-1/\nu} + ...$ for the location of the
maxima of the specific heat
$C = V^{-1}K^2 \left( \langle E^2 \rangle - \langle E \rangle^2 \right)$,
and the connected susceptibility
$\chi^{\rm c} = V K \left( \langle m^2 \rangle -
\langle m \rangle ^2 \right)$.
Using our value of $\nu = 0.704$
we obtain from the linear fits shown in Fig.~1 the estimates
$K_c = 0.6925(9)$ (from $T_{C_{\rm max}}$ with $Q=0.80$) and
$K_c = 0.6930(3)$ (from $T_{\chi^{\rm c}_{\rm max}}$ with $Q=1.0$),
respectively. These values are consistent with the crossing
value (\ref{eq:4}), but have larger statistical errors.
\begin{figure}[htb]
    \vspace{6.8cm}
    \caption{Estimates of the critical temperature $T_c$, coming from Binder's
    crossing method and from the scaling of $T_{C_{\rm max}}$ and
    $T_{\chi^{\rm c}_{\rm max}}$.}
    \label{fig:1}
\end{figure}

The ratio of exponents $\beta/\nu$ follows from the scaling of the
magnetization, $\langle m \rangle \propto L^{-\beta/\nu}$.
{}From the linear least-square fit in a log-log plot of $\langle m \rangle$
versus $L$ at
$K_c = 0.6930$  we obtain the estimate
\begin{equation}
\beta/\nu = 0.514 \pm 0.001,
\label{eq:7}
\end{equation}
(with $Q=0.68$) that is slightly lower than the value given by PFL,
$\beta/\nu = 0.516(3)$
(determined at $K=0.6929$). To test by how much our result is biased
by the value of $K_c$ we have redone our analysis at $K = 0.6929$
($K=0.6931$).
Here we get the slightly higher (lower) value of $0.519(1)$ ($0.509(1)$). The
quality of the fits, however, is much worse, namely $Q=0.30$ ($Q=0.31$),
which we interpret as support for our estimate of $K_c$.
We rely on the goodness-of-fit parameter since visually it
is impossible to make a distinction between these fits when plotted on
a natural scale. It
should be emphasized that even these slight variations in the estimate of
the critical coupling significantly change the estimate of the
exponent ratio in a way that clearly dominates the statistical errors, making
it
necessary to have an accurate estimate of $K_c$.

%
\begin{table*}[hbt]
\setlength{\tabcolsep}{0.80pc}
\newlength{\digitwidth} \settowidth{\digitwidth}{\rm 0}
\catcode`?=\active \def?{\kern\digitwidth}
\caption[a]{Selected sources for $K_c$, $U^*$, and the critical exponents
of the classical 3D Heisenberg model.}
\label{tab:exponents}
  \begin{tabular}{ccccccc}
   \hline
\multicolumn{1}{c}{method}  &
\multicolumn{1}{c}{$K_c$}   &
\multicolumn{1}{c}{$U^*$}   &
\multicolumn{1}{c}{$\nu$}   &
\multicolumn{1}{c}{$\beta / \nu$} &
\multicolumn{1}{c}{$\eta$}   &
\multicolumn{1}{c}{$\alpha / \nu$} \\
\hline
$g$-expansion \cite{pert}                  &
&           & 0.705(3)? & 0.517(6)? & 0.033(4)? & $-0.163(12)$ \\
$\epsilon$-expansion \cite{brezin,epsilon} &
& 0.59684?? & 0.710(7)? & 0.518(11) & 0.040(3)? & $-0.183(28)$ \\
HTS \cite{ahj92}      & 0.6929(1)
&           & 0.712(10) & 0.513(50) & 0.034(42) & $-0.191(40)$ \\
MC \cite{peczak91}    & 0.6929(1)
& 0.622(1)? & 0.706(9)? & 0.516(3)? & 0.031(7)? & $-0.167(36)$ \\
MC (this work)        & 0.6930(1)
& 0.6217(8) & 0.704(6)? & 0.514(1)? & 0.028(2)? & $-0.159(24)$ \\
\hline
\end{tabular}
\end{table*}
Relying on the scaling law $\eta = 2 - \gamma / \nu = 2 \beta / \nu - 1$
we estimate
\begin{equation}
\eta = 0.028 \pm 0.002.
\label{eq:8}
\end{equation}
One can get independent estimates of $\eta$ by direct measurements
of $\chi = VK \langle m^2 \rangle$ and $\chi^{\rm c}$
at our best estimate of $K_c = 0.6930$,
where they both should scale like $L^{ 2 - \eta }$.
{}From linear fits we obtained $\eta = 0.0271(17)$ ($Q=0.78$) and
$\eta = 0.0156(44)$ ($Q=0.69$), respectively. Finally, analyzing the
FSS behavior of the susceptibility maximum,
$\chi^{\rm c}_{\rm max} \propto L^{2 - \eta}$, we estimate $\eta =
0.0231(61)$ ($Q=0.60$). Notice that all MC estimates are lower than the
field theory values, which are collected in Table~1.

Similarly, using the hyper-scaling law  $\alpha/\nu = 2/\nu -3$
we obtain for the specific-heat exponent
\begin{equation}
\alpha/\nu = -0.159 \pm 0.024.
\label{eq:9}
\end{equation}
Measurements of the specific heat are difficult to analyze directly,
because $\alpha$ is negative, which implies
a finite, cusp-like
singularity. We tried a three-parameter fit of the form
$C_{\rm max} = a - b L^{\alpha/\nu}$. The result
$\alpha/\nu = -0.33(22)$ ($Q=0.69$) is compatible with eq.~(\ref{eq:9}),
but has large
statistical errors. Another way of testing eq.~(\ref{eq:9}) is to assume the
predicted value of $\alpha/\nu$ and to fit only the parameters
$a$ and $b$. The resulting fit turned out to be of almost equally good
quality \cite{us}.

To summarize, using the single-cluster
MC update for the classical 3D Heisenberg model on
simple cubic lattices of size up to $48^3$, we obtained high-precision data.
Using multi-histogram techniques we optimally combined the data
and performed a fairly detailed FSS analysis.
Qualitatively, our main result is that the asymptotic FSS region sets
in for small lattices sizes, $L \approx 12$.
Quantitatively, our value for the critical coupling, $K_c = 0.6930(1)$,
is significantly higher than estimates
from old HTS expansion analyses and TM MC methods,
but is in almost perfect agreement with the MC estimate reported recently
by  PFL \cite{peczak91}, with new analyses of longer HTS
expansions \cite{ahj92}, and
with recent MC simulations in the high-temperature phase \cite{hj92}.
Our results for the two basic critical exponents, $\nu = 0.704(6)$ and
$\beta = 0.362(4)$, are in good agreement with field theoretic
predictions.
Direct measurements of $\eta$ give good agreement with the scaling
prediction when the
scaling of $\chi$ at $K_c$ is considered. Using $\chi^{\rm c}$ at
$K_c$ or $\chi^{\rm c}_{\rm max}$, however, the situation is less clear.
In the case of $\alpha/\nu$, its negative value causes numerical problems,
which result in large statistical errors.
Table~1 lists our measured values and their scaling implications
for $\eta$ and $\alpha/\nu$. For comparison, various other
sources for the  critical exponents are added.
More details of this
study can be found in refs. \cite{us,hj92}.
%
%
                 \section*{ACKNOWLEDGEMENTS}
%
We thank the Konrad-Zuse Zentrum Berlin (ZIB) and
the university of Kiel for their generous support
on their CRAY computer resources.
%


\begin{thebibliography}{19}
%
\bibitem{peczak91}
P. Peczak, A.L. Ferrenberg, and D.P. Landau, Phys. Rev. {\bf B43} (1991) 6087.
%
\bibitem{fs}
A.M. Ferrenberg and R.H. Swendsen, Phys. Rev. Lett.
{\bf 61} (1988) 2635; {\em ibid.\/} {\bf 63} (1989) 1195; and
{\em Erratum\/}, {\em ibid.\/} {\bf 63} (1989) 1658.
%
\bibitem{newseries}
S. McKenzie, C. Domb, and D.L. Hunter, J. Phys. {\bf A15} (1982) 3899.
%
\bibitem{bloete}
M.P. Nightingale and H.W.J. Bl\"ote, Phys. Rev. Lett. {\bf 60} (1988) 1562.
%
\bibitem{wolff}
U. Wolff, Phys. Rev. Lett. {\bf 62} (1989) 361.
%
\bibitem{othermodels}
U. Wolff, Phys. Lett. {\bf B228} (1989) 379;\\ 
J.-S. Wang, Physica {\bf A164} (1990) 240;\\
M. Hasenbusch and S. Meyer, Phys. Lett. {\bf B241} (1990) 238;\\
W. Janke, Phys. Lett. {\bf A148} (1990) 306.
%
\bibitem{brezin}
E. Brezin and J. Zinn-Justin, Nucl. Phys. {\bf B257}[FS14] (1985) 867.
%
\bibitem{pert}
J.C. Le Guillou and J. Zinn-Justin, Phys. Rev. {\bf B21} (1980) 3976.
%
\bibitem{epsilon}
J.C. Le Guillou and J. Zinn-Justin, J. Physique Lett. {\bf
46} (1985) L137.
%
\bibitem{us}
For more details, see
C. Holm and W. Janke, preprint FUB-HEP 9/92, HLRZ 56/92, Berlin/J\"ulich
(August 1992).
%
\bibitem{ahj92}
J. Adler, C. Holm, and W. Janke, preprint FUB-HEP 18/92,
HLRZ 76/92 (Sept. 1992).
%
\bibitem{hj92}
C. Holm and W. Janke, preprint FUB-HEP 19/92, HLRZ 77/92 (September 1992).
%

\end{thebibliography}
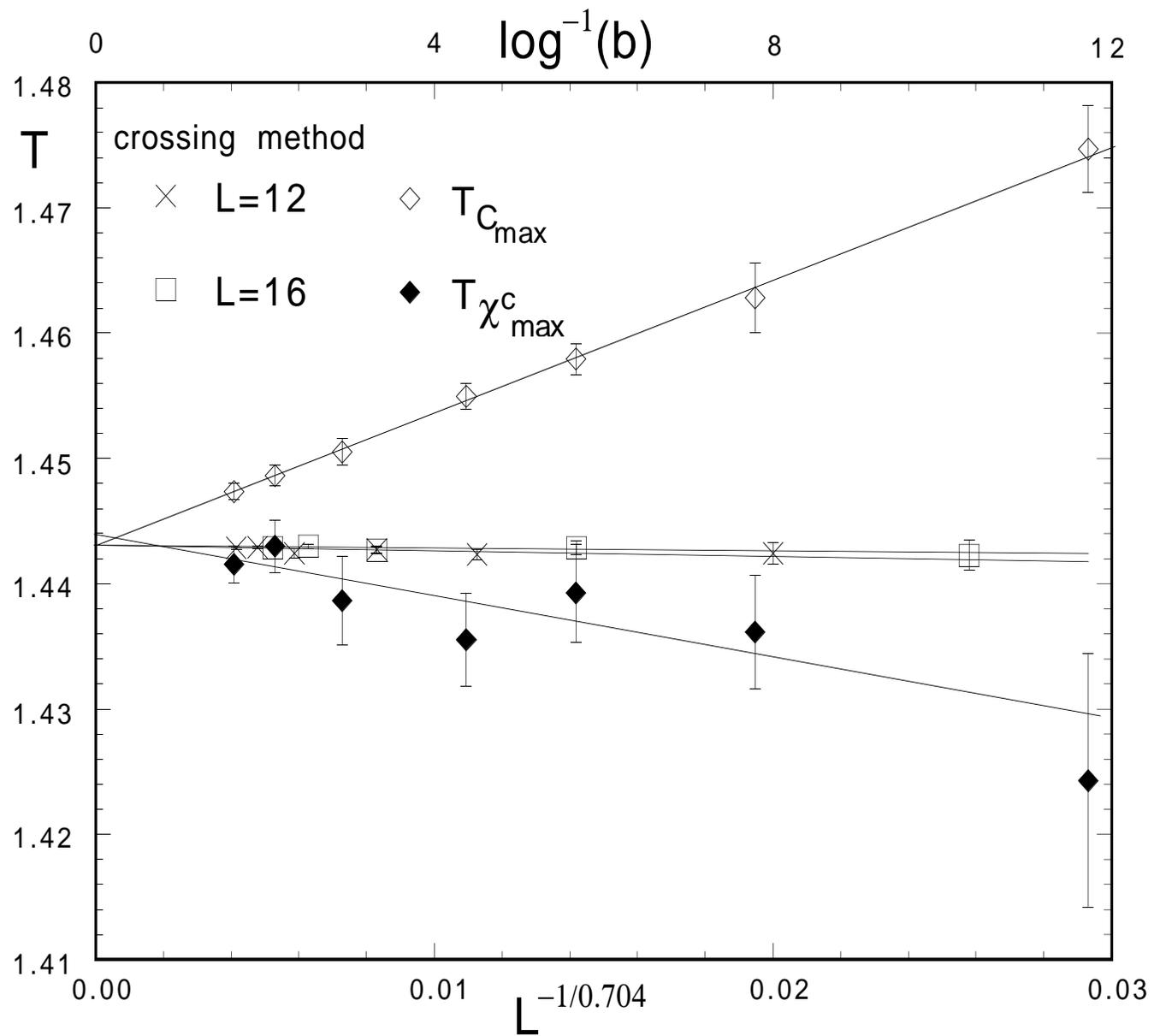
\end{document}